\documentclass[aps,twocolumn,prd,nofootinbib]{revtex4}
\usepackage{amsmath}
\usepackage{graphicx}
\usepackage{dcolumn}
\usepackage{bm}
\usepackage{amssymb}
\usepackage{latexsym}

\begin{document}

\title{Thermodynamics of Spinor Quintom}

\author{Jing Wang$^{1,2}$\footnote{jwang@bao.ac.cn},~~ Shu-wang Cui$^2$,~~Cheng-min Zhang$^1$}
\affiliation{${}$1.National Astronomical Observatories, Chinese
Academy of Sciences, Beijing, China;\\ 2.College of Physics
Science and Information Engineering, \\Hebei Normal University,
Shijiazhuang 050016, P. R. China}

\begin{abstract}

We discuss the thermodynamic properties of dark energy (DE) with
Quintom matter in spinor scenario. (1).Using the Cardy-Verlinde
formula, we investigate the conditions of validity of the
Generalized Second Law of thermodynamics (GSL) in the four
evolutionary phases of Spinor Quintom-B model. We also clarify its
relation with three cosmological entropy bounds. (2). We take
thermodynamic stability of the combination between Spinor Quintom
DE and the generalized Chaplygin Gas (GCG) perfect fluid into
account, and we find that in the case of $\beta>0$ and $0<T<T_0$,
the system we consider is thermodynamically stable. (3) Making use
of the Maxwell Relation and integrability condition, we derive all
thermal quantities as functions of either entropy or volume, and
present the relation with quantum perturbation stability.

\end{abstract}
\maketitle

\section{Introduction}

There are mounting data from type Ia supernovae, cosmic microwave
background (CMB) radiation, and so
on\cite{1998snia,Spergel,Riess,Seljak}, have provided strong
evidences for the present spatially flat and accelerated expanding
universe, corresponding to $\ddot{a}>0$, which is dominated by
dark sectors. Combined analysis of the above cosmological
observations support that the energy of our universe is occupied
by dark energy(DE) about $73\%$, dark matter about $23\%$ and
usual baryon matter only about $4\%$ which can be described by the
well known particle theory. In the context of
Friedmann-Robertson-Walker(FRW) cosmology, the evolution of scale
factor is governed by the temporal part of Enistein equation
$3\frac{\ddot{a}}{a}=-4\pi G(\rho+3p)$, this acceleration may be
attributed to the exotic form of negative pressure satisfying
$p<-3\rho$, the so-called DE. So far, the nature of DE remains a
mystery. To describe the property of this component, a significant
parameter $w=\frac{p}{\rho}$, called Equation of State (EoS), was
introduced. And it is need to be $w<-\frac{1}{3}$ theoretically.
Based on different evolution of the EoS we can obtain different
candidate for DE. Currently, it is widely taken the candidate as a
small cosmological constant $\Lambda$ (or vacuum energy) with EoS
$w=-1$ as well as  a dynamical component such as the Quintessence
with $-1<w<1$\cite{Wetterich:1987fm, Ratra:1987rm}, Phantom with
$w<-1$\cite{Caldwell:1999ew}, K-essence with both $w\geq-1$ and
$w<-1$ but never crossing
$-1$\cite{ArmendarizPicon:2000ah,Chiba:1999ka}. Although the
recent fits to the data in combination of
WMAP\cite{Spergel:2006hy,Komatsu:2008hk}, the recently released
182 SNIa Gold sample\cite{Riess:2006fw} and also other
cosmological observational data show remarkably the consistence of
the cosmological constant, it is worth of noting that a class of
dynamical models with the EoS across $-1$ {\it Quintom} is mildly
favored \cite{Feng:2004ad,Zhao:2006qg, Zhao:2006bt, Wang:2006ts}.
In the literature there have been a lot of theoretical studies of
Quintom-like models. Especially, a No-Go theorem has been proved
to constrain the model building of Quintom\cite{Xia:2007km}, and
according to this No-Go theorem there are models which involve
higher derivative terms for a single scalar field
\cite{Li:2005fm}, models with vector field
\cite{ArmendarizPicon:2004pm}, making use of an extended theory of
gravity \cite{Cai:2005ie}, non-local string field theory
\cite{Aref'eva:2005fu}, and others (see e.g. \cite{Guo:2004fq,
Quintom_tf, Cai:2006dm, Quintom_1, Cai:2007qw, Cai:2007gs,
Cai:2007zv, Quintom_others, Xiong:2007cn}). The similar work
applied in scalar-tensor theory is also studied in Ref.
\cite{Elizalde:2004mq}.

Except that many works have been done in pursuit of establishing
concrete model to understanding the theoretical nature and origin
of this special fluid, there also are a number of papers
committing themselves to investigating the thermodynamic
properties of DE fluid. The thermodynamics of de Sitter space-time
was first investigated by Gibbons and Hawking\cite{Gibbons:1977mu}
and \cite{Verlinde:2000wg, Cai:2009rd, Pollock:1989pn,
Frolov:2002va} extended the study to quasi-de Sitter space-time.
Based on an assumption that DE is a thermallized ensemble at
certain temperature with an associated thermodynamical entropy,
Ref. \cite{Brevik:2004sd, Lima:2004wf,
GonzalezDiaz:2004eu,Izquierdo:2005ku, MohseniSadjadi:2005ps,
Setare:2006vz, Setare:2006rf, Wang:2005pk, Santos:2006ce,
Santos:2007jy, Bilic:2008zk} made various aspects of the
thermodynamic discussions. The
papers\cite{Sheykhi:2008qr,Sheykhi:2008qs} have studied the GSL of
modified gravity. In the literature\cite{Husain:2008qx}, the
thermodynamics of Quantum Gravity has been investigated.
Ref.\cite{Li:2008tc} considered the apparent horizon of the
Friedmann-Robertson-Walker universe as a thermodynamical system
and investigate the thermodynamics of LQC in the semiclassical
region.

Previously, it have been considered that a Quintom dark energy
with non-regular spinor matter\cite{Cai:2008gk}. In succession, to
understand the possible combinations among different types of
Quintom model in spinor field we study the implications of cosmic
duality with this class of models and realize additional Quintom
models by the aid of this dual properties. In the meantime, we
also perform the statefinder diagnostic for this Spinor Quintom
model\cite{wang:2008cs}. In this paper, we will discuss the
thermodynamics of the Spinor Quintom model. From the
thermodynamical point of view, our universe can be considered as a
thermodynamical system filled with DE perfect fluid, we will
examine the GSL and thermodynamic stability in this system. This
letter is organized as follows. In section 2, we investigate the
validity of GSL in spinor field with Quintom DE model, we indicate
that the conditions under which the GSL can be satisfied. In
section 3, we explore the conditions for thermodynamic stability
of the combination between Quintom model with spinor field and the
GCG perfect fluid. Some thermodynamic parameters, as functions of
entropy and volume, are given in section 4, and we also display
the relation with the stability from the point of view of quantum
perturbation stability. Section 5 contains our conclusions and
prospects.

\section{GSL in a System Filled with Spinor Quintom Matter}

One of the distinguishing features of the driver of current
accelerating expansion, the alleged DE, lies in violating the
strong energy condition, $\rho+3p>0$\cite{Riess,Sper}. As a result
of the dependence on theoretical models this strength of
acceleration is a question in debating. While most model
independent analysis suggest that it be below the De Sitter
value\cite{Daly:2003iy}, it is certainly true that the body of
observational data allows of a wide parameter space compatible
with an acceleration larger than de
Sitter's\cite{Caldwell:1999ew,Hannestad:2004cb}. If eventually it
is proven to be the case, this dark component would violate not
only the strong energy condition $\rho+3p>0$ but also the
dominated energy condition $\rho+p>0$. In the literature,
component with the above specialities was dubbed
Phantom\cite{Caldwell:1999ew,Caldwell:2003vq}, suffering from a
long list of pathologies such as quantum
instabilities\cite{Carroll:2003st,Frampton:2003xg} which leads to
supersonic and causes a super accelerating universe ending in a
big rip or big crunch along the cosmic evolution. Attracting many
attentions, the interesting fluid has been widely discussed recent
years\cite{Dabrowski:2003jm,Meng:2003tc}, and
Ref.\cite{GonzalezDiaz:2004eu,Myung:2008km} investigated the
thermodynamics on phantom dark energy dominant universe. The
thermodynamics of DE with constant EoS in the range of
$-1<w<-\frac{1}{3}$ was considered in \cite{Danielsson:2004xw},
and that of K-essence also was studied in Ref.\cite{Bilic:2008zk}.

Based on the relation between the event of horizon and the
thermodynamics of a black hole assumed by Bekenstein in 1973
\cite{Bekenstein:1973ur}, the event of horizon of a black hole is
a measure of its entropy. This idea has been generalized to
horizons of cosmological models, so that each horizon corresponds
to an entropy. Correspondingly, the second law of thermodynamics
was modified in the way that in generalized form, the sum of all
time derivative of entropies related to horizons plus time
derivative of normal entropy must be positive, i.e., the sum of
entropies must be increasing with time. Ref. \cite{Davies:1987ti}
investigated the validity of GSL for the cosmological models which
departs slightly from de Sitter space. Ref.\cite{Izquierdo:2005ku}
explored the thermodynamics of DE taking the existence of the
observer's event horizon in accelerated universes into account.
The conditions of validity of generalized second law in phantom
dominated era was studied in \cite{MohseniSadjadi:2005ps}. The
validity of the GSL of thermodynamics for the Quintom DE model
with two scalar fields without coupling potential term was
considered by \cite{Setare:2006rf}. In this section, we will
discuss the validity of the GSL of thermodynamics for a
Quintom-dominated universe in spinor field and clarify its
relation with three cosmological entropy bounds: the Bekenstein
bound\cite{Bekenstein:1980jp}, the holographic Bekenstein-Hawking
bound, and the Hubble bound\cite{Verlinde:2000wg}.

To begin with the discussion, we deal with the homogeneous and
isotropic Friedmann-Robertson-Walker (FRW) space-time, then the
space-time metric reads,
\begin{equation}
ds^{2}=dt^{2}-a^{2}(t)d\vec{x}^2.
\end{equation}
Assuming that the dynamics of gravity is governed by the
Einstein-Hilbert action, for a spinor minimally coupled to general
relativity\cite{ArmendarizPicon:2003qk,Vakili:2005ya,Ribas:2005vr},
we have,
\begin{equation}
S=S_\psi+S_m-\frac{1}{6}\int d^4 x\sqrt{-g}R.
\end{equation}
where $R$ is the scalar curvature, $S_\psi$ is given by the the
Dirac action and $S_m$ describes additional matter fields, such as
scalar fields and gauge fields.\footnote{Here, we postulate
symmetries, diffeomorphism and local Lorentz invariance.}

We consider the spinor component as the thermodynamical system we
may discuss, which is filled with Quintom DE fluid. With the aid
of the dynamics of a spinor field which is minimally coupled to
Einstein's gravity\cite{Weinberg,BirrellDavies,GSW}, we can write
down the following Dirac action in a curved space-time background
\begin{eqnarray}
S_\psi&=&\int d^4 x~e~[\frac{i}{2}(\bar\psi\Gamma^{\mu}D_{\mu}
\psi-D_{\mu}\bar\psi\Gamma^{\mu}\psi)-\Phi]\nonumber\\
&=&\int d^4 x ~e~{\cal L}_{\psi},
\end{eqnarray}
Here, $e$ is the determinant of the vierbein $e_{\mu}^{a}$ and
$\Phi$ stands for any scalar function of $\psi$, $\bar\psi$ and
possibly additional matter fields. We will assume that $\Phi$ only
depends on the scalar bilinear $\bar\psi\psi$. From the expression
of the Dirac action, we have the energy density and the pressure
of the spinor field:
\begin{eqnarray}
\label{density}\rho_\psi&=&T_{0}^{0}=\Phi~,\\
\label{pressure}p_\psi&=&-T_{i}^{i}=\Phi'\bar\psi\psi-\Phi~,
\end{eqnarray}
For a gauge-transformed homogeneous and a space-independent spinor
field, the equation of motion of spinor reads\cite{Cai:2008gk}
\begin{eqnarray}
\dot{\psi}+\frac{3}{2}H\psi+i\gamma^{0} \Phi' \psi&=&0,\\
\dot{\bar\psi}+\frac{3}{2}H\bar\psi-i\gamma^{0}\Phi' \bar\psi&=&0,
\end{eqnarray}
where a dot denotes a time derivative while a prime denotes a
derivative with respect to $\bar\psi\psi$, and $H$ is Hubble
parameter.

In the framework of FRW cosmology, the Friedmann constraint
equation will be\footnote{Note that we use units $8\pi
G=\hbar=c=1$ and all parameters are normalized by $M_p=1/\sqrt{8
\pi G}$ in the letter.}
\begin{equation}
H^2=\frac{1}{3}\rho_\psi~,
\end{equation}
From the equation of motion of spinor and the Friedmann constraint
equation, we can obtain the the derivative of Hubble parameter
with respect to time,
\begin{equation}
\dot{H}=\frac{\dot{\rho_\psi}}{6H}=\frac{\Phi'\bar{\psi}\psi}{2}.
\end{equation}
So we have
\begin{equation}
\rho_\psi+p_\psi=-2\dot{H}.
\end{equation}

According to the Gibbons equation
\begin{equation}
Tds=dE+p_\psi dV=(p_\psi+\rho_\psi)dV+Vdp_\psi,
\end{equation}
combined with the above relations and the expression of volume
$V=\frac{4}{3}\pi R_H^3$ ($R_H$ is the event of the horizon), we
may rewriting the first law of thermodynamics as,
\begin{eqnarray}
Tds&=&-2\dot{H}d(\frac{4}{3}\pi R_H^3)+\frac{4}{3}\pi R_H^3
d\rho_\psi\nonumber\\
&=&-8\pi R_H^2\dot{H}dR_H+8\pi HR_H^3 dH,
\end{eqnarray}
where $T$ is the temperature of the background of Spinor fluid.
Therefore, the derivative of normal entropy is given as follows:
\begin{equation}
\dot{s}=\frac{ds}{dt}=\frac{1}{T}8\pi\dot{H}R_H^2(HR_H-\dot{R_H}).
\end{equation}
Now we turn to consider the entropy corresponding to the event
horizon. The definition of event horizon in a de Sitter space-time
is
\begin{equation}
R_H=a(t)\int_t^\infty\frac{dt'}{a(t')}.
\end{equation}
So the time derivative of event of horizon in a spinor field
approaching to de Sitter space satisfies the following equation:
\begin{equation}
\dot{R_H}=\dot{a(t)}\int_t^\infty\frac{dt'}{a(t')}+a(t)\dot{\int_t^\infty\frac{dt'}{a(t')}}=HR_H-1.
\end{equation}
i. In the parameter range  of $HR_H\leq1$, the Bekenstein bound,
which is supposed to hold for systems with limited self-gravity,
is appropriate. And the EoS of spinor larger than $-1$,
corresponding to a Quintessence dominant
universe\cite{Davies:1987ti}. ii. While for $HR_H\geq1$,
corresponding to a strongly self-gravitating universe, the
Bekenstein bound has to be replaced by holographic
Bekenstein-Hawking bound in which one has $S_B\geq S_{BH}$. And
one can get a Phantom phase\cite{MohseniSadjadi:2005ps}. iii. If
$HR_H=1$, the Bekenstein bound $S_B$ is equal to holographic
Bekenstein-Hawking bound $S_{BH}$. Then we can write the final
form of the time derivative of normal entropy,
\begin{equation}
\dot{s}=\frac{8\pi R_H^2\dot{H}}{T}.
\end{equation}
As we well know, the entropy is proportional to the area of its
event horizon. If the horizon entropy corresponding to $R_H$ is
defined as $s_H=\pi R_H^2$, the GSL can be stated as:
\begin{equation}
\dot{s}+\dot{s_H}=\frac{8\pi R_H^2\dot{H}}{T}+2\pi
R_H\dot{R_H}\geq0.
\end{equation}

In the following, we will take the Quintom-B model realized in
Ref. \cite{Cai:2008gk} to discuss the validity of GSL in spinor
field. The temperature of Spinor Quintom-B is assumed to be
positive.

(1). Phantom dominated evolution: \\In this phase
$\dot{R_H}\leq0$, so $\dot{s_H}\leq0$. From $V'<0$ one can get
$\dot{H}>0$. So the condition for validity of GSL can be expressed
as:
\begin{equation}
\dot{H}\geq\mid\frac{\dot{R_H}T}{4R_H}\mid.
\end{equation}

(2). Quintessence dominated evolution:\\ In this period of
evolution $\dot{R_H}\geq0$, then we have a negative time
derivative of Hubble parameter but that of horizon entropy is not
a negative value. Thus the condition for validity of GSL is:
\begin{equation}
\mid\dot{H}\mid\leq\frac{\dot{R_H}T}{4R_H}.
\end{equation}

(3). Phase transition from Phantom to Quintessence:\\
At the transition point, we have $w=-1$ and $V'=0$, that is to say
$\dot{H}=0$, so $\dot{s}=0$. Assuming that the event horizon $R_H$
varies continuous, one may expect that $\dot{R_H}=0$ in transition
time, so the horizon entropy is continuous and
differentiable\cite{Setare:2006rf}. Therefore, to realize the
transition, it need to be continuous and differentiable in
transition time for the total entropy of the universe.

(4). The final phase--an approximate de Sitter universe:\\
In such a state, the temperature is \cite{Davies:1987ti},
\begin{equation}
T=\frac{bH}{2\pi},
\end{equation}
where $b$ is a parameter. During this period, the universe lies in
the Quintessence phase, so
\begin{equation}
b\geq\frac{8\pi\mid\dot{H}\mid R_H}{H\dot{R_H}},
\end{equation}
in de Sitter space-time case $R_H=\frac{1}{H}$, one can get
$b\geq8\pi$, which should be satisfied if GSL is valid.

In conclusion, one can find that the conditions for the validity
GSL of Spinor Quintom model are similar to that of the Quintom DE
model constructed by two scalar fields without coupling potential
term which was considered in \cite{Setare:2006rf}.

\section{Thermodynamic Stability of The Combination between Spinor Quintom and GCG Perfect Fluid}

Since the Chaplygin gas was generalized people have make many
correlative studies\cite{Bento:2003dj,Chimento:2003ta} to
reconcile the standard model with observations.
Ref.\cite{Santos:2006ce} discusses the behavior of temperature and
the thermodynamic stability of a generalized Chaplygin gas
considering only general thermodynamics --- the corresponding
thermal equation of state for the GCG and analyzed its temperature
behavior as well as its thermodynamic stability considering both
adiabatic and thermal equations of state. While in the literature
\cite{Santos:2007jy}, Chaplygin gas was modified again, and a
scenario was set up to determine the corresponding thermal
equation of state of the modified Chaplygin gas(MCG) and it
reveals that the MCG only presents thermodynamic stability during
any expansion process if its thermal equation of state depends on
temperature only, $P=P(T)$. Moreover, the modified Chaplygin gas
may cool down through any thermodynamic process without facing any
critical point or phase transition. We have established a
combination between Chaplygin gas and Spinor Quintom in
Ref.\cite{Cai:2008gk}, in this section we will investigate the
thermodynamic stability in a universe filled with the fluid
combined by both Quintom and GCG in spinor field.

In Ref.\cite{Cai:2008gk},we took the form of potential as
$\Phi=\sqrt[1+\beta]{\Phi_0(\bar{\psi}\psi)^{1+\beta}+c}$, and got
the EoS of GCG model
\begin{equation}
p=-\frac{c}{\rho^{\beta}}~,
\end{equation}
where parameter $\beta$ is a constant and positive $\beta>0$ and
$c$ is also positive and a universal constant\cite{Santos:2006ce}.
Here we consider a closed thermodynamic system full of dark energy
fluid, in which the combination of Spinor Quintom with GCG play
important role. Assuming the internal energy $U$ and pressure $p$
as only the functions of their natural viables entropy $s$ and
volume $V$: $U=U(s,V), p=p(s,V)$, and the energy density of DE
fluid is
\begin{equation}
\rho=\frac{U}{V}~.
\end{equation}
From general thermodynamics\cite{Kubo,Landau}, we know that
\begin{equation}
(\frac{\partial U}{\partial V})_s=-p~.
\end{equation}
Combined the above three equations, we can get the following form,
\begin{equation}
(\frac{\partial U}{\partial V})_s=c\frac{V^\beta}{U^\beta},
\end{equation}
and the expression of the internal energy of this system is also
given by its solution,
\begin{equation}
U=\sqrt[1+\beta]{cV^{1+\beta}+b},
\end{equation}
where $b=b(s)$ is an integration parameter. It can be proven that
even $c=c(s)$ is not a universal constant, the above expression
remains valid. The Eq. (25) also can be written
as\cite{Santos:2006ce}:
\begin{equation}
U=V\sqrt[1+\beta]{c[1+(\frac{\sigma}{V})^{1+\beta}]},
\end{equation}
where parameter $\sigma^{1+\beta}=\frac{b}{c}$. Then we may deduce
the expressions of energy density and pressure with respect to
this parameter,
\begin{eqnarray}
\label{density}\rho&=&\sqrt[1+\beta]{c[1+(\frac{\sigma}{V})^{1+\beta}]}~,\\
\label{pressure}p&=&-\sqrt[1+\beta]{\frac{c}{[1+(\frac{\sigma}{V})^{1+\beta}]^{\beta}}}~.
\end{eqnarray}
By these two equations, we could understand the behavior of both
past and future of our universe. In the early time with small
scale factor and volume, the energy density and pressure behave as
the below form:
\begin{eqnarray}
\rho&\approx&c^{\frac{1}{1+\beta}}\frac{\sigma}{V},\\
p&\approx&c^{\frac{1}{1+\beta}}(\frac{V}{\sigma})^{\beta}\sim0,
\end{eqnarray}
corresponding to a high energy density and approximative
pressureless matter dominant phase. During this period the energy
density reduces as its entropy and volume adiabatically. Along
with the cosmological expansion through to some late times, these
two parameters are approximate respectively to
\begin{eqnarray}
\rho&\approx&c^{\frac{1}{1+\beta}}+\frac{c^{\frac{1}{1+\beta}}}{1+\beta}(\frac{\sigma}{V})^{1+\beta},\\
p&\approx&c^{\frac{1}{1+\beta}}.
\end{eqnarray}
During this period of the evolution, the total system can be seen
as constituted by two components: one with constant energy density
and the other with an alterable energy density with respect to
volume. While for a large value of scale factor, the energy
density may rather lower and EoS is
$p=-\rho=c^{\frac{1}{1+\beta}}$ which is a de Sitter Space-time.
Consequently, we realize a transformation from dust-like matter-
dominated universe to a de Sitter phase in the point of view of
thermodynamics.

In what follows, we will extensively examine the conditions for
the thermodynamic stability of this combined system.

(1). We determine how the pressure change with volume through the
adiabatic expansion.\\ Using Eq. (28), one can get
\begin{equation}
(\frac{\partial p}{\partial
V})_s=\beta\frac{p}{V}[1-\frac{1}{1+(\frac{\sigma}{V})^{1+\beta}}],
\end{equation}
it is obvious that we exclude the case of $\beta=0$ due to a
constant pressure and the disappearing derivative. While in the
case of $\beta>0$ the above derivative is always negative value.

(2). To make a system stable, it is necessary for the thermal
capacity at constant volume to be positive $c_V>0$, the pressure
reduces as volume at constant temperature, as well.\\
For this purpose, we calculate the formula of temperature $T$ and
entropy $s$ to determine how the temperature depends on its
entropy and volume. In the thermodynamics and statistical physics,
the temperature of a system is defined as:
\begin{equation}
T=(\frac{\partial U}{\partial s})_V,
\end{equation}
combined with the expression of internal energy, the formula of
temperature can be written as follows\cite{Santos:2006ce}:
\begin{equation}
T=\frac{1}{1+\beta}(cV^{1+\beta}+\varepsilon)^{-\frac{\beta}{1+\beta}}(V^{1+\beta}\frac{dc}{ds}+\frac{d\varepsilon}{ds}).
\end{equation}
Clearly, if we take parameter as both $c$ and $\varepsilon$ are
universal constant, the temperature equals to $0$ for any value of
pressure and volume. As a result, the isotherm $T=0$ is
simultaneously an isentropic curve at $s=const$, which violates
the third law of thermodynamics\cite{Santos:2006ce}. Taking this
factor into account, we choose $c$ as a universal constant and
$\frac{d\varepsilon}{ds}>0$. From dimensional analysis it can be
understood that $\varepsilon$ has a dimension of energy,
$[\varepsilon]^{1+\beta}=[U]$. In this case, we take it
as\cite{Santos:2006ce}
\begin{equation}
b=(T_0s)^{1+\beta},
\end{equation}
so,
\begin{equation}
\frac{d\varepsilon}{ds}=(1+\beta)(T_0s)^{\beta}T_0.
\end{equation}
Then the formulae of temperature and entropy of this system can be
written as:
\begin{eqnarray}
T&=&T_0^{1+\beta}s^{\beta}[cV^{1+\beta}+(T_0s)^{1+\beta}]^{-\frac{\beta}{1+\beta}},\\
s&=&\frac{c^{\frac{1}{1+\beta}}}{T_0}\frac{T^{\frac{1}{\beta}}}{(T_0^{\frac{1+\beta}{\beta}}-T^{\frac{1+\beta}{\beta}})^{\frac{1}{1+\beta}}}V.
\end{eqnarray}
A stable thermodynamic system requires a positive and finite
entropy, which requests that the temperature satisfy
\begin{equation}
0<T<T_0.
\end{equation}
By the definition of $c_V$ and the formulae of temperature and
entropy, one can rewrite $c_V$ as,
\begin{equation}
c_V=\frac{1}{\beta
T_0}\frac{c^{\frac{1}{\beta}}V}{[1-(\frac{T}{T_0})^{\frac{1+\beta}{\beta}}]^{\frac{2+\beta}{1+\beta}}}(\frac{T}{T_0})^{\frac{1}{\beta}},
\end{equation}
Thus, When $\beta>0$ and $0<T<T_0$, one can get a positive $c_V$.

Correspondingly, we can obtain the expression of pressure,
\begin{equation}
p=-c^{\frac{1}{1+\beta}}[1-(\frac{T}{T_0})^{\frac{1+\beta}{\beta}}]^{\frac{\beta}{1+\beta}},
\end{equation}
It can be seen that the pressure is only the function of
temperature, so $(\frac{\partial p}{\partial V})_T>0$ is
satisfied.

In a word, in the case of $\beta>0$ and $0<T<T_0$, the system we
consider is thermodynamically stable.

\section{Thermodynamic parameters and its relation with quantum stabilities}

In the first two sections, we have studied the stability of a
system filled with Spinor Quintom DE fluid from the classical
thermodynamic point of view. For this part, we will derive a class
of thermal quantities as functions either of entropy or volume,
then we may discuss the relation with quantum perturbation and
which constraint is much stronger.

From the expressions of energy density (EQ.\ref{density}) and
pressure (EQ.\ref{pressure}), we can get,
\begin{eqnarray}
&&\rho+P=\sqrt[1+\beta]{c(1+(\frac{\sigma}{V})^{1+\beta})}-\sqrt[1+\beta]{\frac{c}{1+(\frac{\sigma}{V})^{1+\beta}}}\nonumber\\
&&=\sqrt[1+\beta]{c}(\sqrt[1+\beta]{1+(\frac{\sigma}{V})^{1+\beta}}-\frac{1}{\sqrt[1+\beta]{1+(\frac{\sigma}{V})^{1+\beta}}}).
\end{eqnarray}
Besides, From the definition of entropy
\begin{equation}
S\equiv\frac{\rho+P}{T}V,
\end{equation}
we can derive a defining equation of tempertature for an adiabatic
process,
\begin{equation}
\label{T}T\equiv\frac{\rho+P}{S}V.
\end{equation}
Then we have the temperature
\begin{eqnarray}
T_{(V)}&=&\frac{\sqrt[1+\beta]{c}}{S}(\sqrt[1+\beta]{V^{1+\beta}+\sigma^{1+\beta}}\nonumber\\&&-\frac{V^2}{\sqrt[1+\beta]{V^{1+\beta}+\sigma^{1+\beta}}}).
\end{eqnarray}
In addition, the EoS $W_{V}$, squared speed of sound $C^2_{s(V)}$
and entropy $S_{V}$ read respectively,
\begin{eqnarray}
W_{(V)}&=&\frac{P}{\rho}=-\frac{V^{1+\beta}}{V^{1+\beta}+\sigma^{1+\beta}},\\
C^2_{s(V)}&=&\frac{\partial
P}{\partial\rho}=\frac{V^{1+\beta}}{\sigma^{1+\beta}},\\
S_{(V)}&=&\frac{C^{\frac{1}{1+\beta}}}{S}(\sqrt[1+\beta]{V^{1+\beta}+\sigma^{1+\beta}}\nonumber\\&&-\frac{V^2}{\sqrt[1+\beta]{V^{1+\beta}+\sigma^{1+\beta}}}).
\end{eqnarray}

The combination among the integrability condition
\begin{equation}
\frac{\partial^2S}{\partial T\partial
V}=\frac{\partial^2S}{\partial V\partial T},
\end{equation}
the Maxwell Relation
\begin{equation}
\frac{\partial T}{\partial V}=-\frac{\partial P}{\partial S},
\end{equation}
and EQ.(\ref{T}), can lead to the relation,
\begin{equation}
dP=-\frac{\rho+P}{S}dS.
\end{equation}
And setting $\beta=1$ in EQ. (\ref{density}) and EQ.
(\ref{pressure}), one has
\begin{eqnarray}
\rho+P&=&-\sqrt{c}\frac{\frac{\sigma^2}{V^2}}{\sqrt{1+(\frac{\sigma}{V})^2}}
\nonumber\\
&=&\frac{c}{P}-P,
\end{eqnarray}
so
\begin{equation}
\frac{PdP}{P^2-c}=\frac{dS}{S}.
\end{equation}
Finally we can get the thermal quantities as functions of entropy.
\begin{eqnarray}
P_{(S)}&=&-\sqrt{c}\sqrt{1-(\frac{S}{S_*})^2},\\
\rho_(S)&=&\frac{\sqrt{c}}{\sqrt{1-(\frac{S}{S_*})^2}},\\
W_{(S)}&=&(\frac{S}{S_*})^2-1,\\
C^2_{S(S)}&=&1-(\frac{S}{S_*})^2.
\end{eqnarray}
Based on the above expressions of these quantities, we may analyze
the quantum stability in connection with perturbations which is
one important issue of a DE model. Usually systems with negative
kinetic modes from ghost fields suffer from the quantum
instabilities which may induce some supersonic phenomenon.
However, in our Spinor Quintom DE model, we do not introduce any
ghost field, and is it to say that this model will not perform any
quantum instability? To study this issue, we would like to
redefine the spinor as $\psi_N\equiv a^{\frac{3}{2}}\psi$. Then
perturbing the spinor field, one gives the perturbation equation
as follows \cite{Cai:2008gk},
\begin{eqnarray}\label{perteq}
&&\frac{d^2}{d\tau^2}\delta\psi_N-\nabla^2\delta\psi_N+\nonumber\\
&&a^2\left[ V'^2+i\gamma^0 (HV'-3HV''\bar\psi\psi)
\right]\delta\psi_N\nonumber\\
&&=-2a^2V'V''\delta(\bar\psi\psi)\psi_N\nonumber\\
&&-i\gamma^\mu\partial_\mu[a V''\delta(\bar\psi\psi)]\psi_N~,
\end{eqnarray}
where $\tau$ is the conformal time defined by $d\tau\equiv dt/a$.
From the perturbation equation above, we can read that the sound
speed is equal to $1$ which eliminates the instability of the
system in short wavelength.

Thus to what degree the system is stable in both quantum and
classical level, and which constraint is much stronger.
Furthermore, whether there are some instability from the
unrenormalizable quantum effect. Such issues we may discuss in
detail in our future work.

\section{Conclusion and Discussions}

To summarize, we have investigated the thermodynamics of Quintom
DE dominant thermodynamical system in spinor field. Firstly, we
show the conditions in which the total entropy may not decrease
with time not only in Phantom and Quintessence phase but also at
the transition time and the final approximative de Sitter phase.
We set up the similar conditions to a Quintom universe with two
scalar fields without coupling potential term. In the second
place, we, using general thermodynamics, explore the thermodynamic
stability of a system full of the DE fluid combined Spinor Quintom
with GCG, and we conclude that in a certain range of temperature,
i.t. $0<T<T_0$, this system remains thermodynamically stable
without any limitation on pressure. We also derive a class of
thermal quantities as functions either of entropy or volume, then
we may discuss the relation with quantum perturbation. And in our
future work, we may clarify which constraint is much stronger by
detailed calculations.

\section*{Acknowledgements}
It is a pleasure to thank Xin-min Zhang for helpful discussions
and advisements. This work is supported in part by Natural
Sciences Foundation of China (Nos. 10975046), NSFC (No.10773017)
and National Basic Research Program of China (2009CB824800).


\vfill

\begin{thebibliography}{99}

\bibitem{1998snia}
  S. Perlmutter {\it et al.},
  Astrophys. J. {\bf 483}, 565 (1997);
  Adam G. Riess et al.,
  Astrophys. J. {\bf116}, 1009 (1998).

\bibitem{Spergel}
  D. N. Spergel {\it et al.},
  Astrophys. J. Suppl. {\bf 148}, 175 (2003).

\bibitem{Riess}
  A. G. Riess {\it et al.},
  Astrophys. J. {\bf 607}, 665 (2004).

\bibitem{Seljak}
  U.~Seljak {\it et al.},
  Phys.\ Rev.\  D {\bf 71}, 103515 (2005)
  [arXiv:astro-ph/0407372].


\bibitem{Wetterich:1987fm}
  C.~Wetterich,
  Nucl.\ Phys.\  B {\bf 302}, 668 (1988).

\bibitem{Ratra:1987rm}
  B.~Ratra and P.~J.~E.~Peebles,
  Phys.\ Rev.\  D {\bf 37}, 3406 (1988).


\bibitem{Caldwell:1999ew}
  R.~R.~Caldwell,
  Phys.\ Lett.\  B {\bf 545}, 23 (2002)
  [arXiv:astro-ph/9908168].


\bibitem{ArmendarizPicon:2000ah}
  C.~Armendariz-Picon, V.~F.~Mukhanov and P.~J.~Steinhardt,
  Phys.\ Rev.\  D {\bf 63}, 103510 (2001)
  [arXiv:astro-ph/0006373].

\bibitem{Chiba:1999ka}
  T.~Chiba, T.~Okabe and M.~Yamaguchi,
  Phys.\ Rev.\  D {\bf 62}, 023511 (2000)
  [arXiv:astro-ph/9912463].

\bibitem{Spergel:2006hy}
  D.~N.~Spergel {\it et al.}  [WMAP Collaboration],
  Astrophys.\ J.\ Suppl.\  {\bf 170}, 377 (2007)
  [arXiv:astro-ph/0603449].

\bibitem{Komatsu:2008hk}
  E.~Komatsu {\it et al.}  [WMAP Collaboration],
  arXiv:0803.0547 [astro-ph].


\bibitem{Riess:2006fw}
  A.~G.~Riess {\it et al.},
  arXiv:astro-ph/0611572.

\bibitem{Feng:2004ad}
  B.~Feng, X.~L.~Wang and X.~M.~Zhang,
  Phys.\ Lett.\  B {\bf 607}, 35 (2005)
  [arXiv:astro-ph/0404224].

\bibitem{Zhao:2006qg}
  G.~B.~Zhao, J.~Q.~Xia, H.~Li, C.~Tao, J.~M.~Virey, Z.~H.~Zhu and X.~Zhang,
  Phys.\ Lett.\  B {\bf 648}, 8 (2007)
  [arXiv:astro-ph/0612728].

\bibitem{Zhao:2006bt}
  G.~B.~Zhao, J.~Q.~Xia, B.~Feng and X.~Zhang,
  Int.\ J.\ Mod.\ Phys.\  D {\bf 16}, 1229 (2007)
  [arXiv:astro-ph/0603621].

\bibitem{Wang:2006ts}
  Y.~Wang and P.~Mukherjee,
  Astrophys.\ J.\  {\bf 650}, 1 (2006)
  [arXiv:astro-ph/0604051].


\bibitem{Xia:2007km}
  J.~Q.~Xia, Y.~F.~Cai, T.~T.~Qiu, G.~B.~Zhao and X.~Zhang,
  Int.\ J.\ Mod.\ Phys.\  D {\bf 17}, 1229 (2008)
  [arXiv:astro-ph/0703202].

\bibitem{Li:2005fm}
  M.~Z.~Li, B.~Feng and X.~M.~Zhang,
  JCAP {\bf 0512}, 002 (2005)
  [arXiv:hep-ph/0503268].

\bibitem{ArmendarizPicon:2004pm}
  C.~Armendariz-Picon,
  JCAP {\bf 0407}, 007 (2004)
  [arXiv:astro-ph/0405267];
  H.~Wei and R.~G.~Cai,
  Phys.\ Rev.\  D {\bf 73}, 083002 (2006)
  [arXiv:astro-ph/0603052].

\bibitem{Cai:2005ie}
  R.~G.~Cai, H.~S.~Zhang and A.~Wang,
  Commun.\ Theor.\ Phys.\  {\bf 44}, 948 (2005)
  [arXiv:hep-th/0505186];
  P.~S.~Apostolopoulos and N.~Tetradis,
  Phys.\ Rev.\  D {\bf 74}, 064021 (2006)
  [arXiv:hep-th/0604014];
  K.~Bamba, C.~Q.~Geng, S.~Nojiri and S.~D.~Odintsov,
  arXiv:0810.4296 [hep-th].

\bibitem{Aref'eva:2005fu}
  I.~Y.~Aref'eva, A.~S.~Koshelev and S.~Y.~Vernov,
  Phys.\ Rev.\  D {\bf 72}, 064017 (2005)
  [arXiv:astro-ph/0507067];
  S.~Y.~Vernov,
  arXiv:astro-ph/0612487;
  A.~S.~Koshelev,
  JHEP {\bf 0704}, 029 (2007)
  [arXiv:hep-th/0701103];

\bibitem{Guo:2004fq}
  Z.~K.~Guo, Y.~S.~Piao, X.~M.~Zhang and Y.~Z.~Zhang,
  Phys.\ Lett.\ {\bf B608}, 177 (2005)
  [arXiv:astro-ph/0410654].

\bibitem{Quintom_tf}
  X.~F.~Zhang, H.~Li, Y.~S.~Piao and X.~M.~Zhang,
  Mod.\ Phys.\ Lett.\  A {\bf 21}, 231 (2006)
  [arXiv:astro-ph/0501652];
  Z.~K.~Guo, Y.~S.~Piao, X.~Zhang and Y.~Z.~Zhang,
  Phys.\ Rev.\  D {\bf 74}, 127304 (2006)
  [arXiv:astro-ph/0608165].


\bibitem{Cai:2006dm}
  Y.~F.~Cai, H.~Li, Y.~S.~Piao and X.~M.~Zhang,
  Phys.\ Lett.\  B {\bf 646}, 141 (2007)
  [arXiv:gr-qc/0609039].


\bibitem{Quintom_1}
  B.~Feng, M.~Li, Y.~S.~Piao and X.~Zhang,
  Phys.\ Lett.\  B {\bf 634}, 101 (2006)
  [arXiv:astro-ph/0407432];
  H.~Wei and R.~G.~Cai,
  Phys.\ Rev.\  D {\bf 72}, 123507 (2005)
  [arXiv:astro-ph/0509328];
  X.~Zhang,
  Phys.\ Rev.\  D {\bf 74}, 103505 (2006)
  [arXiv:astro-ph/0609699].

\bibitem{Cai:2007qw}
  Y.~F.~Cai, T.~Qiu, Y.~S.~Piao, M.~Li and X.~Zhang,
  JHEP {\bf 0710}, 071 (2007)
  [arXiv:0704.1090 [gr-qc]].

\bibitem{Cai:2007gs}
  Y.~F.~Cai, M.~Z.~Li, J.~X.~Lu, Y.~S.~Piao, T.~T.~Qiu and X.~M.~Zhang,
  Phys.\ Lett.\  B {\bf 651}, 1 (2007)
  [arXiv:hep-th/0701016].

\bibitem{Cai:2007zv}
  quintom is found to be able to give a bouncing solution and
  its perturbations combine aspects of singular and nonsingular
  bounce models, see for example:
  Y.~F.~Cai, T.~Qiu, R.~Brandenberger, Y.~S.~Piao and X.~Zhang,
  JCAP {\bf 0803}, 013 (2008)
  [arXiv:0711.2187 [hep-th]];
  Y.~F.~Cai, T.~T.~Qiu, J.~Q.~Xia and X.~Zhang,
  arXiv:0808.0819 [astro-ph];
  Y.~F.~Cai and X.~Zhang,
  arXiv:0808.2551 [astro-ph].

\bibitem{Quintom_others}
  W.~Zhao, and Y. Zhang,
  Phys.\ Rev.\  D {\bf 73}, 123509 (2006)
  [arXiv:astro-ph/0604460];
  H.~Mohseni Sadjadi and M.~Alimohammadi,
  Phys.\ Rev.\  D {\bf 74}, 043506 (2006)
  [arXiv:gr-qc/0605143];
  E.~O.~Kahya and V.~K.~Onemli,
  Phys.\ Rev.\  D {\bf 76}, 043512 (2007)
  [arXiv:gr-qc/0612026];
  Y.~F.~Cai and Y.~S.~Piao,
  Phys.\ Lett.\  B {\bf 657}, 1 (2007)
  [arXiv:gr-qc/0701114].
  R.~Lazkoz, G.~Leon and I.~Quiros,
  Phys.\ Lett.\  B {\bf 649}, 103 (2007)
  [arXiv:astro-ph/0701353];
  H.~Zhang and Z.~H.~Zhu,
  arXiv:0704.3121 [astro-ph];
  T.~Qiu, Y.~F.~Cai and X.~M.~Zhang,
  arXiv:0710.0115 [gr-qc];
  M.~R.~Setare, J.~Sadeghi and A.~Banijamali,
  Phys.\ Lett.\  B {\bf 669}, 9 (2008)
  [arXiv:0807.0077 [hep-th]].

\bibitem{Xiong:2007cn}
  H.~H.~Xiong, T.~Qiu, Y.~F.~Cai and X.~Zhang,
  arXiv:0711.4469 [hep-th];
  H.~H.~Xiong, Y.~F.~Cai, T.~Qiu, Y.~S.~Piao and X.~Zhang,
  Phys.\ Lett.\  B {\bf 666}, 212 (2008)
  [arXiv:0805.0413 [astro-ph]];
  S.~Li, Y.~F.~Cai and Y.~S.~Piao,
  arXiv:0806.2363 [hep-ph];
  S.~Zhang and B.~Chen,
  Phys.\ Lett.\  B {\bf 669}, 4 (2008)
  [arXiv:0806.4435 [hep-ph]].


\bibitem{Elizalde:2004mq}
  E.~Elizalde, S.~Nojiri and S.~D.~Odintsov,
  Phys.\ Rev.\  D {\bf 70}, 043539 (2004)
  [arXiv:hep-th/0405034].


\bibitem{Gibbons:1977mu}
  G.~W.~Gibbons and S.~W.~Hawking,
  Phys.\ Rev.\  D {\bf 15}, 2738 (1977).

\bibitem{Verlinde:2000wg}
  E.~P.~Verlinde,
  arXiv:hep-th/0008140.

\bibitem{Cai:2009rd}
  Y.~F.~Cai, W.~Xue, R.~Brandenberger and X.~m.~Zhang,
  JCAP {\bf 0906}, 037 (2009)
  [arXiv:0903.4938 [hep-th]].

\bibitem{Pollock:1989pn}
  M.~D.~Pollock and T.~P.~Singh,
  Class.\ Quant.\ Grav.\  {\bf 6} (1989) 901.

\bibitem{Frolov:2002va}
  A.~V.~Frolov and L.~Kofman,
  JCAP {\bf 0305}, 009 (2003)
  [arXiv:hep-th/0212327].


\bibitem{Brevik:2004sd}
  I.~Brevik, S.~Nojiri, S.~D.~Odintsov and L.~Vanzo,
  Phys.\ Rev.\  D {\bf 70}, 043520 (2004)
  [arXiv:hep-th/0401073].
  S.~Nojiri and S.~D.~Odintsov,
  Phys.\ Rev.\  D {\bf 70}, 103522 (2004)
  [arXiv:hep-th/0408170].

\bibitem{Lima:2004wf}
  J.~A.~S.~Lima and J.~S.~Alcaniz,
  Phys.\ Lett.\  B {\bf 600}, 191 (2004)
  [arXiv:astro-ph/0402265].

\bibitem{GonzalezDiaz:2004eu}
  P.~F.~Gonzalez-Diaz and C.~L.~Siguenza,
  Nucl.\ Phys.\  B {\bf 697}, 363 (2004)
  [arXiv:astro-ph/0407421].
  S.~H.~Pereira and J.~A.~S.~Lima,
  arXiv:0806.0682 [astro-ph].

\bibitem{Izquierdo:2005ku}
  G.~Izquierdo and D.~Pavon,
  Phys.\ Lett.\  B {\bf 633}, 420 (2006)
  [arXiv:astro-ph/0505601].

\bibitem{MohseniSadjadi:2005ps}
  H.~Mohseni Sadjadi,
  Phys.\ Rev.\  D {\bf 73}, 063525 (2006)
  [arXiv:gr-qc/0512140].

\bibitem{Setare:2006vz}
  M.~R.~Setare and S.~Shafei,
  JCAP {\bf 0609}, 011 (2006)
  [arXiv:gr-qc/0606103].

\bibitem{Setare:2006rf}
  M.~R.~Setare,
  Phys.\ Lett.\  B {\bf 641}, 130 (2006)
  [arXiv:hep-th/0611165].

\bibitem{Wang:2005pk}
  B.~Wang, Y.~Gong and E.~Abdalla,
  Phys.\ Rev.\  D {\bf 74}, 083520 (2006)
  [arXiv:gr-qc/0511051].
  Y.~Gong, B.~Wang and A.~Wang,
  Phys.\ Rev.\  D {\bf 75}, 123516 (2007)
  [arXiv:gr-qc/0611155].
  B.~Wang, C.~Y.~Lin, D.~Pavon and E.~Abdalla,
  Phys.\ Lett.\  B {\bf 662}, 1 (2008)
  [arXiv:0711.2214 [hep-th]].


\bibitem{Santos:2006ce}
  F.~C.~Santos, M.~L.~Bedran and V.~Soares,
  Phys.\ Lett.\  B {\bf 636} (2006) 86.

\bibitem{Santos:2007jy}
  F.~C.~Santos, V.~Soares and M.~L.~Bedran,
  Phys.\ Lett.\  B {\bf 646}, 215 (2007).

\bibitem{Bilic:2008zk}
  N.~Bilic,
  arXiv:0806.0642 [gr-qc].

\bibitem{Sheykhi:2008qr}
  A.~Sheykhi and B.~Wang,
  arXiv:0811.4477 [hep-th].

\bibitem{Sheykhi:2008qs}
  A.~Sheykhi and B.~Wang,
  arXiv:0811.4478 [hep-th].


\bibitem{Husain:2008qx}
  V.~Husain and R.~B.~Mann,
  arXiv:0812.0399 [gr-qc].

\bibitem{Li:2008tc}
  L.~F.~Li and J.~Y.~Zhu,
  arXiv:0812.3544 [gr-qc].


\bibitem{Cai:2008gk}
  Y.~F.~Cai and J.~Wang,
  Class.\ Quant.\ Grav.\  {\bf 25}, 165014 (2008)
  [arXiv:0806.3890 [hep-th]].

\bibitem{wang:2008cs}
Jing Wang and Shi-ping Yang,
(submit to Phys. Lett. B)

\bibitem{Sper}
S.perlmutter,et,al.,  Nature, {\bf391}, 51 (1998)

\bibitem{Daly:2003iy}
  R.~A.~Daly and S.~G.~Djorgovski,
  Astrophys.\ J.\  {\bf 597}, 9 (2003)
  [arXiv:astro-ph/0305197].
  M.~V.~John,
  Astrophys.\ J.\  {\bf 614}, 1 (2004)
  [arXiv:astro-ph/0406444].
  S.~Nesseris and L.~Perivolaropoulos,
  Phys.\ Rev.\  D {\bf 70}, 043531 (2004)
  [arXiv:astro-ph/0401556].
  Y.~Wang and M.~Tegmark,
  Phys.\ Rev.\  D {\bf 71}, 103513 (2005)
  [arXiv:astro-ph/0501351].

\bibitem{Hannestad:2004cb}
  S.~Hannestad and E.~Mortsell,
  JCAP {\bf 0409}, 001 (2004)
  [arXiv:astro-ph/0407259].
  J.~Q.~Xia, G.~B.~Zhao, B.~Feng, H.~Li and X.~Zhang,
  Phys.\ Rev.\  D {\bf 73}, 063521 (2006)
  [arXiv:astro-ph/0511625].

\bibitem{Caldwell:2003vq}
  R.~R.~Caldwell, M.~Kamionkowski and N.~N.~Weinberg,
  Phys.\ Rev.\ Lett.\  {\bf 91}, 071301 (2003)
  [arXiv:astro-ph/0302506].

\bibitem{Carroll:2003st}
  S.~M.~Carroll, M.~Hoffman and M.~Trodden,
  Phys.\ Rev.\  D {\bf 68}, 023509 (2003)
  [arXiv:astro-ph/0301273].
  J.~M.~Cline, S.~Jeon and G.~D.~Moore,
  Phys.\ Rev.\  D {\bf 70}, 043543 (2004)
  [arXiv:hep-ph/0311312].

\bibitem{Frampton:2003xg}
  P.~H.~Frampton,
  Mod.\ Phys.\ Lett.\  A {\bf 19}, 801 (2004)
  [arXiv:hep-th/0302007].

\bibitem{Dabrowski:2003jm}
  M.~P.~Dabrowski, T.~Stachowiak and M.~Szydlowski,
  Phys.\ Rev.\  D {\bf 68}, 103519 (2003)
  [arXiv:hep-th/0307128].
  G.~W.~Gibbons,
  arXiv:hep-th/0302199.
  A.~E.~Schulz and M.~J.~White,
  Phys.\ Rev.\  D {\bf 64}, 043514 (2001)
  [arXiv:astro-ph/0104112].
  P.~Singh, M.~Sami and N.~Dadhich,
  Phys.\ Rev.\  D {\bf 68}, 023522 (2003)
  [arXiv:hep-th/0305110].
  Z.~K.~Guo, Y.~S.~Piao and Y.~Z.~Zhang,
  Phys.\ Lett.\  B {\bf 594}, 247 (2004)
  [arXiv:astro-ph/0404225].

\bibitem{Meng:2003tc}
  X.~H.~Meng and P.~Wang,
  arXiv:hep-ph/0311070.
  V.~B.~Johri,
  Phys.\ Rev.\  D {\bf 70}, 041303 (2004)
  [arXiv:astro-ph/0311293].
  L.~P.~Chimento and R.~Lazkoz,
  Phys.\ Rev.\ Lett.\  {\bf 91}, 211301 (2003)
  [arXiv:gr-qc/0307111].
  L.~P.~Chimento and R.~Lazkoz,
  Phys.\ Rev.\ Lett.\  {\bf 91}, 211301 (2003)
  [arXiv:gr-qc/0307111].
  L.~P.~Chimento and D.~Pavon,
  Phys.\ Rev.\  D {\bf 73}, 063511 (2006)
  [arXiv:gr-qc/0505096].
  M.~Sami and A.~Toporensky,
  Mod.\ Phys.\ Lett.\  A {\bf 19}, 1509 (2004)
  [arXiv:gr-qc/0312009].
  M.~Szydlowski, W.~Czaja and A.~Krawiec,
  Phys.\ Rev.\  E {\bf 72}, 036221 (2005)
  [arXiv:astro-ph/0401293].
  M.~Bouhmadi-Lopez and J.~A.~Jimenez Madrid,
  JCAP {\bf 0505}, 005 (2005)
  [arXiv:astro-ph/0404540].
  Y.~H.~Wei and Y.~Tian,
  Class.\ Quant.\ Grav.\  {\bf 21}, 5347 (2004)
  [arXiv:gr-qc/0405038].
  V.~K.~Onemli and R.~P.~Woodard,
  Phys.\ Rev.\  D {\bf 70}, 107301 (2004)
  [arXiv:gr-qc/0406098].
  P.~F.~Gonzalez-Diaz,
  TSPU Vestnik {\bf 44N7}, 36 (2004)
  [arXiv:hep-th/0408225].

\bibitem{Myung:2008km}
  Y.~S.~Myung,
  arXiv:0810.4385 [gr-qc].

\bibitem{Danielsson:2004xw}
  U.~H.~Danielsson,
  Phys.\ Rev.\  D {\bf 71}, 023516 (2005)
  [arXiv:hep-th/0411172].

\bibitem{Bekenstein:1973ur}
  J.~D.~Bekenstein,
  Phys.\ Rev.\  D {\bf 7} (1973) 2333.

\bibitem{Davies:1987ti}
  P.~C.~W.~Davies,
  Class.\ Quant.\ Grav.\  {\bf 4}, L225 (1987).

\bibitem{ArmendarizPicon:2003qk}
  C.~Armendariz-Picon and P.~B.~Greene,
  Gen.\ Rel.\ Grav.\  {\bf 35}, 1637 (2003)
  [arXiv:hep-th/0301129].

\bibitem{Vakili:2005ya}
  B.~Vakili, S.~Jalalzadeh and H.~R.~Sepangi,
  JCAP {\bf 0505}, 006 (2005)
  [arXiv:gr-qc/0502076].

\bibitem{Ribas:2005vr}
  M.~O.~Ribas, F.~P.~Devecchi and G.~M.~Kremer,
  Phys.\ Rev.\  D {\bf 72}, 123502 (2005)
  [arXiv:gr-qc/0511099].

\bibitem{Weinberg}
  S. Weinberg, \textit{Gravitation and Cosmology}, Cambridge University Press (1972).

\bibitem{BirrellDavies}
  N. Birrell and P. Davies, \textit{Quantum Fields in Curved Space}, Cambridge University Press (1982).

\bibitem{GSW}
  M. Green, J.  Schwarz, E.   Witten, \textit{Superstring Theory Vol. 2}, Chapter 12, Cambridge University Press (1987).

\bibitem{Bekenstein:1980jp}
  J.~D.~Bekenstein,
  Phys.\ Rev.\  D {\bf 23}, 287 (1981).
  J.~D.~Bekenstein,
  Phys.\ Rev.\  D {\bf 49}, 1912 (1994)
  [arXiv:gr-qc/9307035].
  J.~D.~Bekenstein,
  Int.\ J.\ Theor.\ Phys.\  {\bf 28}, 967 (1989).

\bibitem{Bento:2003dj}
  A.~Dev, D.~Jain and J.~S.~Alcaniz,
  Phys.\ Rev.\  D {\bf 67}, 023515 (2003)
  [arXiv:astro-ph/0209379].
  M.~C.~Bento, O.~Bertolami and A.~A.~Sen,
  Gen.\ Rel.\ Grav.\  {\bf 35}, 2063 (2003)
  [arXiv:gr-qc/0305086].
  L.~Amendola, F.~Finelli, C.~Burigana and D.~Carturan,
  JCAP {\bf 0307}, 005 (2003)
  [arXiv:astro-ph/0304325].
  R.~Bean and O.~Dore,
  Phys.\ Rev.\  D {\bf 68}, 023515 (2003)
  [arXiv:astro-ph/0301308].
  M.~Szydlowski and W.~Czaja,
  Phys.\ Rev.\  D {\bf 69}, 023506 (2004)
  [arXiv:astro-ph/0306579].
  T.~Multamaki, M.~Manera and E.~Gaztanaga,
  Phys.\ Rev.\  D {\bf 69}, 023004 (2004)
  [arXiv:astro-ph/0307533].

\bibitem{Chimento:2003ta}
  L.~P.~Chimento,
  Phys.\ Rev.\  D {\bf 69}, 123517 (2004)
  [arXiv:astro-ph/0311613].

\bibitem{Kubo}
  R.Kubo, Thermodynamics, North-Holland, Amsterdam, 1968.

\bibitem{Landau}
  L.D.Landau, E.M.Lifschitz,  Statitical Physics, third, ed.,
  Course of Theoretical Physics, vol.5, Butterworth-Heinemann,
  London, 1984.






\end{thebibliography}
\end{document}